# Diffusive Shock Acceleration Simulations: Comparison with Particle Methods and Bow Shock Measurements

Hyesung Kang

Department of Earth Sciences, Pusan National University, Pusan 609-735, Korea; e-mail: kang@astrophys.es.pusan.ac.kr

and

T. W. Jones

Department of Astronomy, University of Minnesota, Minneapolis, MN 55455; e-mail: twj@astro.spa.umn.edu

## ABSTRACT

Direct comparisons of diffusive particle acceleration numerical simulations have been made against Monte Carlo and hybrid plasma simulations by Ellison *et. al.* (1993) and against observations at the earth's bow shock presented by Ellison *et. al.* (1990). Toward this end we have introduced a new numerical scheme for injection of cosmic-ray particles out of the thermal plasma, modeled by way of the diffusive scattering process itself; that is, the diffusion and acceleration across the shock front of particles out of the suprathermal tail of the Maxwellian distribution. Our simulations take two forms. First, we have solved numerically the time dependent diffusion-advection equation for the high energy (cosmic-ray) protons in one-dimensional quasi-parallel shocks. Dynamical feedback between the particles and thermal plasma is included. The proton fluxes on both sides of the shock derived from our method are consistent with those calculated by Ellison *et. al.* (1993). A similar test has compared our methods to published measurements at the earth's bow shock when the interplanetary magnetic field was almost parallel to the solar wind velocity (Ellison *et. al.* 1990). Again our results are in good agreement.

Second, the same shock conditions have been simulated with the two-fluid version of diffusive shock acceleration theory by adopting injection rates and the closure parameters inferred from the diffusion-advection equation calculations. The acceleration efficiency and the shock structure calculated with the two-fluid method are in good agreement with those computed with the diffusion-advection method.

Thus, we find that all of these computational methods (diffusion-advection, two-fluid, Monte Carlo and hybrid) are in substantial agreement on the issues they can simultaneously address, so that the essential physics of diffusive particle acceleration is adequately contained within each. This is despite the fact that each makes what appear to be very different assumptions or approximations.

*Subject headings:* Cosmic-Rays— particle acceleration— hydrodynamics

## 1. Introduction

Strong collisionless shocks in astrophysical environments are among the most widely discussed sites for the acceleration of energetic charged particles (Cosmic-Rays, CR hereafter). The first order Fermi process, by which particles reach high energies incrementally through multiple shock crossings, gained special favor from the late 1970s with the introduction of the "diffusive shock acceleration" model (Axford et al. 1977, Bell 1978, Blandford & Eichler 1987). That model has since been broadly accepted, refined and examined (see reviews by Drury 1983, Blandford & Eichler 1987, Berezhko & Krymskii 1988, Jones & Ellison 1991). Although simple in concept, the full problem of diffusive shock acceleration is actually extremely complex, because the nonlinear interactions between energetic particles, resonantly scattering waves and the underlying plasma have to be considered. Important consequences of nonlinear interactions deriving from the presence of CR include such things as generation and damping of the scattering wave field, as well as heating and compression of the plasma flow due to the CR pressure. Such subtleties reflect the fact that the physics of collisionless shocks is itself very complex (e.g., Kennel, Edmiston & Hada 1985).

Several distinct approaches have developed to probe the important issues associated with the production of high energy particles at collisionless shocks. The diffusive shock acceleration model focuses on the CRs as a relatively small population of high energy particles distinguished from the "thermal" plasma by larger particle scattering lengths expected to accompany larger energies. Here particle scattering length means the distance over which a particle loses all memory of its initial pitch angle through collisionless, collective, electromagnetic interactions associated with underlying magnetic irregularities. While "thermal" particles are presumed to be scattered within a relatively thin "subshock", CRs are presumed to have scattering lengths long enough to allow them relatively free passage through the subshock. CRs, however, are still supposed to have their momenta randomized on finite length scales, so that their motion through the background medium is approximately diffusive on the macroscopic scales of interest. Diffusive acceleration itself is included through Fokker-Planck scattering terms in a kinetic equation for the CR particle distribution. The kinetic equation for CRs ought to be solved simultaneously with a set of fluid equations describing the flow associated with the bulk, thermal plasma, including the nonlinear interactions between the plasma, CRs and scattering waves. In the past this was a difficult computational challenge, but it is now beginning to be practically possible to compute fairly complete, nonlinear versions of diffusive shock acceleration models and apply them to real, time dependent, multidimensional astrophysical problems.

On the other hand, separating out the CR population from thermal gas is somewhat artificial,

especially for energies near those of thermal particles. Indeed, it has become recognized that particles from the thermal plasma can be "injected" at shocks into the more energetic CR population (e.g., Eichler 1979, Quest 1988). Also, the fluid treatment of the shock transition itself is often idealized to a discontinuity (or the numerical approximation thereof), whereas real collisionless shocks are not so simple. The detailed microphysics of collisionless shock formation and the acceleration of thermal particles to suprathermal energies are very complex and beyond the scope of this study.

One particular issue of special relevance to particle acceleration is the feature that the suprathermal particles involved in "injection" processes have intermediate scattering lengths, but seem also to exhibit substantial streaming motions in the vicinity of the shock (e.g., Scholar 1990). They cannot be modeled accurately by either the gasdynamics equations or the CR kinetic equation. Since they eschew self-consistent treatment in a straightforward way within the framework of a diffusive shock acceleration/fluid dynamic formalism, one has to devise a special macroscopic model that contains the essential physics of the processes that influence them. A variety of approaches have been attempted. Injection models have usually been based on the idea that once their momenta are large enough particles should behave more like CRs than thermal particles. Generally they have depended on simple parameterizations, such as fixing a fraction of the thermal proton flux through the shock to be injected at some fiducial suprathermal momentum (*e.g.*, Bell 1987, Falle & Giddings 1987, Kang & Jones 1991). Two-fluid calculations have used extensions of this (Kang & Jones 1990) or similar models based on injection of a small, fixed fraction of the energy flux (Markiewicz et al. 1990, Dorfi 1991). But, these methods are unavoidably *ad hoc* to some degree, as discussed, for example, by Zank, Webb & Donohue 1993. It would be better to design an injection model that more closely reflects such characterizations as "thermal leakage" (Ellison & Eichler 1984). In such a model particles within the thermal pool that were accelerated above a threshold momentum through normal, collective "plasma" processes would be more properly handled through finite scale scattering than as part of a "collisionally dominated" fluid.

Proper injection would include acknowledgment of the fact that it must depend on the form of the distribution function, $f(p)$, especially beyond the Maxwellian peak. As a step in that direction Zank, Webb & Donohue 1993 suggested a "thermal leakage" type injection model for two-fluid CR transport in which particles with momenta greater than an injection momentum $p_i$ contribute to CR energy and pressure while particles less energetic than $p_i$ are included in the thermal plasma. They pointed out that thermal particles would leak into the CR distribution by crossing $p_i$ as a consequence of adiabatic flow compression. Thus, the particle distribution $f(p_i)$, along with the gradient of the flow velocity, which establishes the rate of momentum change in a converging flow, determines the injection rate. However, two-fluid models follow CR energy and pressure, but do not track the actual particle distribution, so this still requires one to work with a free parameter. They introduced $\alpha = (4\pi/3)(\frac{1}{2}mp_i^2)p_i^3 f(p_i)$, which we cannot know without a full solution to the problem, so it is still an assumed quantity in practice.

Issues associated with self-consistent shock structure and injection physics are more directly addressed through so-called "plasma simulation" or through Monte Carlo simulation techniques (Quest 1988, Scholar 1990, Giacalone, et al. 1993, Ellison & Eichler 1984, Ellison et al. 1990). For simplicity we will henceforth call such approaches "particle methods". They follow individual particle motions and, in principle, allow a full determination of the entire particle distribution as well as the shock structure itself, sometimes including the electromagnetic wave fields. But, particle methods do not easily lend themselves to problems involving complex geometries and extended, time dependent flows, so they are not as readily applied to some "large-scale" astrophysical problems. In this context we introduce below a "thermal leakage" injection model for the diffusive kinetic equation method in which the particle distribution in the tail beyond the Maxwellian peak is treated separately from both thermal gas and CRs. The particle injection rate is modeled through "numerically controlled" diffusive acceleration of these intermediate momentum particles into CRs in an effort to mimic the macroscopic consequences of a combination of the microphysics of shock formation and the fact that some particles become excited to energies high enough to represent CR particle injection. The results of diffusive acceleration simulations with the new injection model are compared directly with solutions from particle methods in the present paper.

Since particle methods and diffusive transport methods begin from very different perspectives, such comparisons between them should be extremely useful to provide firm grounds for applying each of the methods where they work best and for establishing any practical limitations imposed by the approximations made in each case. Comparison between nonlinear calculations and measured properties of real shocks would be even more illuminating. Some recent work has begun similar evaluations in earnest with regard to particle methods. Ellison et al. 1990 (hereafter EMP) have compared Monte Carlo simulations of quasi-parallel shocks with particle measurements taken from the Earth's bow shock. They demonstrated that Monte Carlo methods can simulate the energy dissipation in the shock and also the injection and acceleration of particles in a manner consistent with observation. Subsequently, Ellison et al. 1993 (hereafter EGBS) successfully compared Monte Carlo and one dimensional hybrid plasma simulations for a model shock rather similar to the real one discussed in EMP. We refer to their papers and citations therein for details on their computational methods. As explained by EGBS, the basic difference in their two procedures is that "the hybrid code follows many particles simultaneously and computes particle trajectories from self-consistently determined magnetic and electric fields, while the Monte Carlo code assumes that a specific scattering law determines the particle motion and that collective effects are modeled only, on average, by the iterative determination of the shock structure and compression ratio." Thus, the methods differ significantly from one another and also in fundamental ways from the diffusive transport methods we apply here. In the EGBS test both particle methods produced similar shock structures and particle distribution functions, demonstrating basic consistency of the two methods. Our intention in the present paper is to extend simulation of the two shock examples from EGBS and EMP to include diffusive shock acceleration models, thus providing the remaining model comparisons for quasi-parallel CR shocks.

In our tests we will include two versions of the diffusive transport model. The original and more general model describes the CRs through their kinetic equation, including a momentum dependent diffusion coefficient, $\kappa(p)$, (equation 2-10). We shall outline that model, which for simplicity we term the "kinetic equation" model, in the next section. However, the kinetic equation model can be computationally very expensive to work with, especially for time dependent simulations, so diffusive acceleration has often been studied through a simplified hydrodynamic, "two-fluid" model. In the two-fluid model CRs are represented as a massless fluid that interacts with the thermal plasma through an isotropic pressure, $P_c$, and is characterized by an adiabatic index, $\gamma_c$, (Drury & Völk 1981). Two-fluid CR diffusion is represented by a momentum averaged diffusion coefficient, $\langle\kappa\rangle$ that accounts for CR energy diffusion, (equation 2-13). This simpler, but much more economical model has been widely used for analytic steady state and time dependent computer simulations to study the nonlinear feedback of the CR pressure on flow dynamics and the acceleration efficiency. Especially it has been applied recently in various time dependent, multidimensional problems (Ryu, Kang & Jones 1993, Jones & Kang 1993, Jones, Kang & Tregillis 1994 ) and models of SNR blast waves (Dorfi 1990, Dorfi 1991, Jones & Kang 1990, Jones & Kang 1992, Drury et al. 1989, Markiewicz et al. 1990, Kang & Drury 1992) in order to investigate the possibility that diffusive acceleration in SNRs is efficient enough to replenish the galactic CRs.

The most serious limitations of the two-fluid model are imposed by the fact that it requires *a priori* estimates of the particle distribution properties in order to specify the closure parameters $\gamma_c$ and $\langle\kappa\rangle$, along with estimates of the energetics associated with shock injection of suprathermal particles into the CR population. This means, of course, that two-fluid methods are not suitable for determining the CR distribution itself. In addition, time-dependent numerical simulations using the two-fluid model have shown that quantitative calculations of diffusive shock acceleration with it can sometimes be rather sensitive to the closure parameters and the assumed injection rate (Achterberg, Blandford, & Periwal 1984, Jones & Kang 1990, Dorfi 1990, Kang & Drury 1992). On the other hand, provided we understand these limitations, so that we apply two-fluid models appropriately to dynamical issues only and adopt suitable models for the closure parameters, the two-fluid model would seem to be a valuable tool to explore nonlinear dynamical effects and the dependence of the acceleration efficiency on the closure parameters and the injection model (see, for example, Jones & Kang 1992, Duffy et al. 1994, Frank, Jones & Ryu 1995).

However, the apparent simplicity of the two-fluid model and concerns over the need to work with *a priori* closure parameters has led to some strongly skeptical views and concerns regarding its suitability even under those constraints (*e.g.,* Achterberg, Blandford, & Periwal 1984, Jones & Ellison 1991). Beyond the specific points already mentioned, several other criticisms have been raised. For example, it is sometimes argued that momentum dependent diffusion cannot be included in two-fluid models, and that two-fluid models cannot properly account for the differing scales on which CR particles of various energies will react with the plasma. Questions have also been raised about consequences of the fact that particle numbers are not explicitly accounted for in two-fluid models and the possibility that this precludes proper inclusion of particle inertia or the effects of the high energy particles escaping from a shock (Jones & Ellison 1991). However, in defense of the two-fluid model one should note that there is nothing fundamental about that model that prevents one from assuming momentum dependent diffusion (see equation 2-13). Rather, the problem is that one must assume what that dependence is. Not only two-fluid models, but also kinetic equation models generally ignore CR inertia, because the expectation is that the inertia should be a small effect compared to the effects of $P_c$ on the gas flow. Further, it is entirely possible to account within the two-fluid model for the dynamical influence of particle escape, as we shall demonstrate below. Again, the criticism should be directed not at fundamental flaws in the two-fluid model, but rather at the issues of whether particle escape is dynamically significant in a particular application and, if so, *how* the effect is included. The importance of those concerns is as likely to apply to kinetic equation simulations as to two-fluid simulations. In this paper we will provide direct comparisons between time dependent kinetic equation simulations and two-fluid simulations in which all of these issues are important, excepting foreknowledge of $\gamma_c$. Duffy et al. 1994 and Frank, Jones & Ryu 1995 have recently demonstrated consistency between kinetic equation and two-fluid simulations when $\gamma_c$ is modeled in appropriate ways.

In the following section we will outline our methods, including our injection model and the basic equations for both the kinetic equation version of diffusive shock acceleration and the simpler two-fluid version. In §3 we discuss our tests of these methods against the earlier published results. Section 4 contains our conclusions.

## 2. Model Description

As discussed above, the diffusive transport model for CR acceleration separates the plasma into two components distinguished by scattering length, generally determined by particle momentum. We treat the bulk plasma, representing lower energy thermal particles whose scattering lengths are small enough to fit within a dissipative shock as an ordinary magnetized fluid. The CR population is followed by a kinetic equation with diffusive terms. The inertia in the CR population is generally neglected in such treatments, and we will do the same. Our treatments of the underlying fluid and the CRs are relatively standard, so we begin this section by describing our new method of modeling the behavior of particles associated with the injection from thermal to CR populations.

### 2.1. A Numerical Injection Model

We emphasized before that diffusive transport models cannot accurately treat the behavior of particles involved directly in injection. The detailed physics of this process is apparently complex and certainly not yet well understood. However, we can identify some features that seem to be important. Bulk energy dissipation and particle scattering in quasi parallel shocks involve collective interactions mediated through electromagnetic wave fields. The structures are neither steady nor simple and seem to involve various streaming or "firehose" instabilities, e.g., Quest 1988, Zachary, etal 1989. Thus, for particles directly involved in mediation of the shock it is hard to characterize accurately their behaviors by any simple rule. The net result seems to be that most particles are reasonably "thermalized", but that a small population end up with excess energy and form a suprathermal "tail" on an otherwise quasi-Maxwellian distribution. The latter are the seed particles we identify as participating in CR injection. In a statistical modeling approach, such as we are employing, it is not practical to follow the short term behavior of wave fields and shock structure. We can take into account that most of the particles participate *en bulk*, forming a structure of finite thickness related to the ion inertial length or the gyroradius of thermalized ions, but that there is a population of particles that seem to take more efficient advantage of the shock energy reservoir by accessing energy near to but upstream of the shock (i.e., the seed particles) (Scholar 1990). Those particles may not be properly described by an isotropic distribution following a diffusion-advection equation, but that is probably a better characterization of them than as part of a thermal distribution confined within the shock. Just as for the CR themselves, a diffusive treatment relies on knowing the spectrum of scattering Alfvén waves. We should properly determine that self-consistently, as discussed, for example, by Jones 1993, Jones 1994. For the present, however, our aim is to compare results to particle simulations that used an assumed scattering behavior. So we defer the more complete treatment.

Our new treatment utilizes the entire particle distribution, $f(p)$, over the range of particle momenta ($p_o \leq p \leq p_3$) that includes the thermal plasma. This last detail is one departure from previous diffusive transport calculations. Below a certain momentum, $p_1$, chosen high enough to include most of the postshock thermal population (and, hence, virtually all of the upstream population), the distribution is forced to maintain a Maxwellian form consistent with the gas density and pressure determined from the gasdynamical equations (see below); *i.e.*, the temperature is $T = \mu(u_1/c)^2(P_g/\rho)$, where for our simulations $\mu = 1$ downstream of the subshock and $\mu = \frac{1}{2}$ upstream, as explained in §2.5. Henceforth we will express particle momenta in units $mc$, so that temperature is most naturally expressed in units $mc^2/k$. Above $p_1$ particles are allowed to evolve according to the diffusion-convection equation, so the form will deviate from Maxwellian. However, only for $p \geq p_2 > p_1$ are they included in calculations of CR pressure and energy. Thus particles are "injected" into the CR population at a lower momentum boundary $p_2$ with an effective injection rate

$$Q(p_2) = 4\pi p_2^2 f(p_2) \left(\frac{dp}{dt}\right)_{p_2}, \qquad (2\text{-}1)$$

where

$$\left(\frac{dp}{dt}\right)_{p_2} = -\frac{1}{3} p_2 \vec{\nabla} \cdot (\vec{u} + \vec{u}_w) \qquad (2\text{-}2)$$

measures the rate of momentum gain through compression in the flow. This leads to an "injection"

energy loss term for the thermal gas,

$$- S = - \left( m_p c^2 \frac{p_2^2}{2} \right) Q(p_2). \tag{2-3}$$

Equation 2-3, using equation 2-1, is the same as that suggested by Zank, Webb & Donohue 1993 for their two-fluid model. It is also contained within a general expression for two-fluid energy source terms given by Kang & Jones 1990. The energy loss term $S$ is calculated numerically and explicitly included in the gas energy equation presented in §2.3, while the injection rate $Q$ is implicitly accounted for through the advection of particles through momentum space at $p_2$. Without the term $S$ in the gas energy equation, of course, total energy is not conserved and the flow becomes much more compressive, as in a non-adiabatic shock such as a radiatively cooling shock.

The particle population in the interval $p_1 < p \leq p_2$ is a kind of virtual injection pool. In that momentum range particle transport is allowed to obey the diffusion-advection equation, so some diffusion takes place. However, with a strong momentum dependence to the diffusion coefficient, energy gains are dominated by adiabatic effects. As mentioned before, the particle distribution in this pool cannot be correct, because the diffusion-advection equation is not valid for the momentum range and the numerical shock structure is not the real shock structure. Thus we can think of $p_1$ as a free parameter that controls the shape of $f$ near the injection momentum $p_2$ allowing the resulting injection rate to be controlled in a manner more analogous to the real processes than simpler models can do.

The choice of $p_1$ influences the injection rate directly in two ways. First, since $f(p_1)$ itself is part of the Maxwellian distribution, it is determined by the local temperature. If we relate $p_1$ to the peak of the Maxwellian distribution, $p_{th}$, as $p_1 = c_1 p_{th}$, with $c_1 > 1$, then one expects injection to increase as $c_1$ is decreased towards unity. Here $p_{th} = 2\sqrt{T}$ corresponds to the peak in the partial pressure of thermal particles $(dP/dp \propto g(p) \propto \left(\frac{1}{T}\right)^{\frac{3}{2}} \exp\left(-\frac{p^2}{2T}\right)$, where $g = p^4 f(p)$.) Second, since particles with $p > p_1$ are supposed to be able to leak upstream from the shock, they should have scattering lengths that exceed the shock thickness. Using forms defined by EGBS and EMP, we will introduce in §2.4 scattering laws for these simulations that have momentum dependencies, $\lambda \propto p^\alpha$, with $\alpha > 0$, and values that exceed the gyroradii, $r_g$, of particles with momenta, $p \sim p_{th}$, by about an order of magnitude. Thus, as discussed further in the §2.5, we have selected numerical grid resolutions with zone widths comparable to $r_g(p_{th})$, so the numerical shock widths are comparable to the scattering lengths of thermal particles, but distinctly less than those for particles in the injection pool or the CR population. Our comparison tests with the shocks being modeled here imply appropriate values $1.5 < (p_1/p_{th}) < 2.0$ under these conditions. By contrast to $p_1$ it is most convenient for the particular test problems here to fix the value of $p_2$ as $p_2 = 3.26 \times 10^{-3}$ (i.e., 5 KeV). The particle distributions shown by EMP and EGBS show a significant deviation from the thermal distribution above this energy and they consider the particle roughly above this energy as CRs. Unlike $p_1$, which ought to be tied to the shock thickness, it is more appropriate to fix the value of $p_2$ rather than the ratio of $p_2/p_{th}$, since $p_{th}$ can decrease significantly as the subshock weakens in response to CR feedback. We find that 3-4 $\times p_{th}$ of the initial postshock gas is a reasonable choice for $p_2$. The dynamics should not be sensitively dependent on the choice of $p_2$ once the partial pressure of CR particles is dominated by particles above $p_2$, which happens long before the shock reaches a dynamical equilibrium in our simulations. Thus we consider $p_2$ as a physically chosen quantity rather than as a free parameter.

To see more clearly these relationships examine Fig. 1. It shows the time evolution of the distribution ($g = p^4 f$) of the protons downstream of the EGBS test shock computed with our methods ($c_1 = 1.87$ and $c_1 = 1.95$). The dashed vertical line represents the momentum $p_2$ at 5 KeV. The position of $p_1$, which is determined by the postshock temperature, changes with time and corresponds to the point above which the distribution function deviates from a Maxwellian distribution in Fig.1. The particles between $p_1$ and $p_2$ diffuse across the shock and are "injected" into CR population when their momenta become greater than $p_2$. One might see this scheme as another injection model with an artificial free parameter. The virtues of this model, however, lie in the fact that the value of $p_1$ can apparently be confined into a reasonably small range by comparison with other methods, and also that the evolution of the particle distribution inside the injection pool more closely approximates real physical processes than some simpler injection models adopted in diffusive acceleration calculations. In addition, it answers the complaint sometimes raised about models that allow injection only at the subshock. Since the procedure we outline takes place everywhere in the grid, injection is possible throughout. However, in practice it is only very close to the subshock that the population in the injection pool becomes significant.

To account for injection in the two-fluid model one needs to transfer energy from the gas to the CR by some prescribed rate in the shock. We can use equation 2-3 only approximately, since we cannot know $f(p_2)$ explicitly, except by using the results of the kinetic equation simulations. In the kinetic equation simulations, we calculate numerically the injection energy rate $S$ and subtract it from the thermal energy. In the two-fluid calculations, however, we determine from those results the spatially integrated injection rate, $I = \int S dx$, previously defined by Kang & Jones 1990 and subtract this energy from several zones around the subshock, since most of the injection takes place around the subshock. Then we can write

$$I = \int S dx = \frac{1}{2} v_2^2 \rho_1 u_1 \int \left[ \frac{\frac{4\pi}{3} p_2^3 f(p_2)}{n_1 u_1} \left( -\frac{du}{dx} \right) \right] dx = \left(\frac{1}{2} v_2^2 \rho_1 u_1\right) \eta, \qquad (2\text{-}4)$$

where $\eta$ is a dimensionless two-fluid "injection parameter", $v_2 = p_2 c$, and $n_1$ is the total proton number density upstream of the shock. We can then calculate $\eta$ from equation 2-4 as a function of time for each shock modeled using the results of the kinetic equation calculations. Assuming that the injection is spread over several numerical zones with a weighting distribution $w$, we can determine a practical two-fluid version of the source term, $S_{tf}$, as in Kang & Jones 1990; namely,

$$S_{tf} = \eta \frac{\left(\frac{1}{2} v_2^2 \rho_1 u_1\right) w}{\Delta x}. \qquad (2\text{-}5)$$

In Kang & Jones 1990 we defined our injection model by causing a fixed fraction, $\epsilon$, of the incident proton flux, $n_1 u_1$, to be injected at particle speed, $\lambda_i c_{s2}$, where $c_{s2}$ is the postshock sound speed and $\lambda_i$ is a parameter we fixed at 2. Provided we numerically distribute the injection around the subshock in the same way in both models, we can formally relate $\eta$ and $\epsilon$ by the expression

$$\eta = \epsilon \left( \lambda_i^2 \frac{c_{s2}^2}{v_2^2} \right). \tag{2-6}$$

For the simulations in this paper, in fact, $v_2 \sim (3-5)\, c_{s2}$, so that $\eta \sim (\frac{1}{2} - \frac{1}{5})\, \epsilon$; i.e., $\eta$ is roughly proportional to the fraction of the incident proton flux that is injected into the CR population at the shock. In practical terms then, the two-fluid version of our new injection model is almost the same as the one we have used before. For our two fluid simulations we distributed $S_{tf}$ with a Gaussian form four zones either side of a numerically flagged subshock zone. In both the EGBS and EMP cases we found from the kinetic equation simulations that $\eta$ quickly increased with time from zero (since $f(p_2)$ was typically very small upstream of the shock at the beginning) to a limiting, constant value. This increase was much faster than the time required to achieve dynamical equilibrium, since there need only be a modest amount of diffusive acceleration between $p_1$ and $p_2$ for $\eta$ to emerge. We carried out two-fluid model simulations allowing for this evolution as well as just assuming a constant $\eta$. Except at the very beginning, there was no significant difference in the results, and certainly none in the time-asymptotic behavior. For the EGBS two-fluid simulation we assumed $\eta = 0.014$, while for the EMP simulation we assumed $\eta = 0.04$.

## 2.2. Basic Dynamical Equations

For quasi-parallel shocks in which the magnetic field plays no role in the bulk gas dynamics, the background fluid follows the equations of ideal gas dynamics generalized to include CR pressure as follows,

$$\frac{d\rho}{dt} = -\rho \vec{\nabla} \cdot \vec{u}, \tag{2-7}$$

$$\frac{d\vec{u}}{dt} = -\frac{1}{\rho}\vec{\nabla}(P_g + P_c), \tag{2-8}$$

$$\frac{de}{dt} = -\frac{1}{\rho}\vec{\nabla} \cdot \{(P_g + P_c)\vec{u}\} + \frac{1}{\rho}P_c\vec{\nabla} \cdot \vec{u} - \frac{S - L}{\rho}. \tag{2-9}$$

Here, $P_g$ and $P_c$ are the pressure contributed from gas and CR particles, respectively, while $\rho$ is the mass density, $\vec{u}$ is the gas flow velocity and $e$ is the sum of gas thermal and kinetic energy per unit mass. The time derivative is the total, Lagrangian derivative measured in the frame of the fluid. The injection energy term $S$ is given by equation 2-3 for the kinetic equation calculations, while it should be replaced with the two-fluid version $S_{tf}$ given by equation 2-5 for two-fluid calculations. $L$ is a heating term that will be used to represent the dissipation of Alfvén waves generated by streaming CRs. It will be discussed in more detail below. For our present purposes we will assume pressure from the resonantly scattering Alfvén waves can be neglected. Flows are aligned in the $\hat{x}$ direction.

Diffusive transport is treated through the particle kinetic equation. The CR population is assumed to be sufficiently strongly scattered both upstream and downstream of the shock to insure that the distribution function $f(x, p, t)$ is isotropic to first order in the reference frame of the scattering centers. The scattering centers are Alfvén waves, which we can allow to propagate at a velocity $\vec{u}_w$ with respect to the fluid frame. For these calculations we will make the simplifying assumption that $\vec{u}_w = \pm \vec{v}_A$, where $\vec{v}_A$ is the local Alfvén velocity pointed in the direction of the local large scale magnetic field. To further simplify the present issues, we will also assume that the important scattering centers are propagating in the upstream direction, since those are the waves that should be most strongly stimulated by CRs in the vicinity of the shock (*e.g.*, Völk, Drury & McKenzie 1984). (The shocks we will construct are right-facing, so this means in practice we will assume $\vec{u}_w = v_A \hat{x}$.) Also we ignore the pressure from He and heavy nuclei, since we expect they contribute only about 10 % of total pressure for either thermal plasma or CR and their exclusion should not alter the form of the proton distribution significantly. Even though EMP and EGBS included these elements and presented the particle flux for them, incorporating them in our calculations would add considerable complexity inappropriate for the level of comparison we intend to make.

Under the above circumstances CR transport can be described by a diffusion-advection equation (Parker 1965, Skilling 1975),

$$\frac{df}{dt} = \frac{1}{3}\vec{\nabla} \cdot (\vec{u} + \vec{u}_w) p \frac{\partial f}{\partial p} + \vec{\nabla} \cdot (\kappa \vec{\nabla} f) - \vec{u}_w \cdot \nabla f, \tag{2-10}$$

where $f(x, p, t)$ is measured in the convected frame, $\vec{u} + \vec{u}_w$, while $d/dt$ is again the total time derivative in the fluid frame. As mentioned above we express momentum, $p$, in units of $mc (= 9.38 \times 10^5 \mathrm{keV}/c$ for the protons). In addition the distribution function $f$ is in units of the particle number density, so that $4\pi \int f p^2 \, dp = \rho/m$. Injection or escape of CRs is implicitly accounted for by appropriate boundary conditions.

The spatial diffusion coefficient, $\kappa(x, p, t)$, is conveniently written in terms of a scattering length, $\lambda$, as

$$\kappa = \frac{1}{3}\lambda v = \frac{1}{3}\frac{\lambda \, pc}{\sqrt{p^2 + 1}}, \tag{2-11}$$

where $v$ is the particle speed. The scattering length, $\lambda$, and thus $\kappa$, are determined by the intensity of resonantly interacting Alfvén waves. Therefore, a fully self-consistent theory must include the evolution of those waves, too (e.g., Jones 1993, Jones 1994). However, since our purpose here is to compare directly with previous particle simulations that provided models for $\lambda$, we will adopt those forms in this paper. To be strictly valid, equation 2-10 requires that the particle speeds be much greater than the flow speed and that the first order anisotropy correction to the distribution function is due to particle streaming in response to spatial gradients in $f$. Momentum diffusion (second order Fermi acceleration) is neglected. The calculations we describe below consider momenta in a range $p_0 \leq p \leq p_3$ sufficiently broad to include both thermal particles and CRs, but as described in §2.1 we count particles in the CR population only when they lie in the more restricted range $p_2 \leq p \leq p_3$, where equation 2-10 should be valid. We recall, as well, that for $p_o < p < p_1$, with $p_1 > p_{th}$, $f(p)$ is forced to remain Maxwellian. So, in effect, equation 2-10 is ignored over this momentum range.

In the two-fluid version of the diffusive acceleration model the energy moment of the diffusion-advection equation is computed from $p_2$ to $p_3$ to produce the conservation equation for CR energy; namely,

$$\frac{dE_c}{dt} = -\gamma_c E_c (\vec{\nabla} \cdot \vec{u}) + \vec{\nabla} \cdot (\langle \kappa \rangle \vec{\nabla} E_c - \vec{u}_w \gamma_c E_c) + \vec{u}_w \cdot \vec{\nabla} P_c + S_{tf}, \qquad (2\text{-}12)$$

where $E_c$ is the CR energy density. No new approximations are introduced in deriving equation 2-12 from equation 2-10, but it does contain three closure parameters, $\gamma_c$, $\langle \kappa \rangle$ and $S_{tf}$ that are really properties of the solution, but in practice must be estimated *a priori*. The injection energy rate, $S_{tf}$ is given by equation 2-5, as before, to represent energy exchange with the thermal plasma. The upper momentum bound, $p_3$ is set high enough that energy flux crossing that boundary can be ignored. However, there will be an implicit source term applied to equation 2-12 through boundary conditions intended to account for particles escaping upstream, as discussed in §2.4. Because of that total energy is not conserved in the present calculations.

The mean diffusion coefficient, $\langle \kappa \rangle$, is defined according to

$$\langle \kappa \rangle = \frac{\int_{p_2}^{p_3} \frac{\partial f}{\partial x} \kappa [\sqrt{p^2+1} - 1] p^2 dp}{\int_{p_2}^{p_3} \frac{\partial f}{\partial x} [\sqrt{p^2+1} - 1] p^2 dp}. \qquad (2\text{-}13)$$

The CR adiabatic index, $\gamma_c$, is defined as

$$\gamma_c = 1 + \frac{P_c}{E_c}, \qquad (2\text{-}14)$$

while

$$P_c = \frac{4\pi m c^2}{3} \int_{p_2}^{p_3} \frac{p^4}{\sqrt{p^2+1}} f \, dp, \qquad (2\text{-}15)$$

and

$$E_c = 4\pi m c^2 \int_{p_2}^{p_3} p^2 (\sqrt{p^2+1} - 1) f \, dp. \qquad (2\text{-}16)$$

### 2.3. Alfvén Wave Transport

Since the Alfvén and sonic Mach numbers are comparable for both the EGBS and EMP shocks, we have included the effects of "Alfvén wave transport" (*e.g.*, Völk, Drury & McKenzie 1984, Jones 1993) in some of the simulations presented here. This allows for the fact that in diffusive transport theory the scattering centers are attached to Alfvén wave "turbulence" rather than the inertial frame of the fluid, so that the propagation speed $\vec{u} + \vec{u}_w$ determines the rate at which CR are advected through the shock. In addition, through the same resonant scattering that isotropizes the CR there is an energy exchange between the CR and the waves that depends on $\vec{u}_w \cdot \vec{\nabla} P_c$ ( cf. equation 2-12). The latter effect leads to wave growth. If that wave growth is balanced by damping in the plasma (Völk, Drury & McKenzie 1984) then this also produces a heating source term for the gas. In the local equilibrium limit this is

$$L = -\vec{u}_w \cdot \vec{\nabla} P_c. \tag{2-17}$$

Jones 1993 emphasized the importance of including both wave advection and dissipation in modeling Alfvén wave transport. For our simulations we have adopted the simple model introduced by Völk, Drury & McKenzie 1984. It assumes $\vec{u}_w = v_A \hat{x}$; *i.e.*, Alfvén waves are generated by CR streaming in the upwind direction. Inclusion of these effects, which can be substantial, heats the gas above the adiabatic rate as it passes through the precursor and also slows the rate of CR acceleration. Consequently, we found in those models that a slightly smaller value of $c_1$ is needed to get similar compression ratios compared to the case without those terms (See Jones 1993 for a discussion of the properties of variations on this model for Alfvén wave transport.)

Our simulations that include Alfvén wave transport assume for the upstream flows that $v_A = c_s$ in the both tests, where $c_s = \sqrt{\gamma_g P_g / \rho}$. Elsewhere in the grid, $v_A \propto \rho^{-\frac{1}{2}}$. The measured magnetic field reported by EMP would yield an upstream $v_A = 2c_s$ in this case. We did carry out simulations with that condition, but were unsuccessful in matching them to the observed particle fluxes and density jump. For such a large Alfvén speed the consequences of Alfvén wave transport are very sensitive to details of the model, which is itself simplified. Therefore, we believe the smaller wave speed is as appropriate a value here to represent the influence of Alfvén wave transport as the larger value.

## 2.4. Model Characteristics

Our simulations are designed for comparison with the results presented by EMP and EGBS. Since the EGBS comparison is somewhat simpler, we introduce that one first. Those authors compared two kinds of simulations of a quasi-parallel collisionless shock. They used both a one-dimensional hybrid plasma simulation and a Monte Carlo method to examine a shock whose basic assumed properties are identified as EGBS and listed in the first three columns of Table 1 . The upstream proton density, $n_1$, temperature, $T_1$, and flow speed with respect to the shock, $u_1$, are given. In addition the Alfvén Mach number in the hybrid simulation was 6.3, or approximately the same as the sonic Mach number. EGBS presented results computed by both particle methods for the distribution functions upstream and downstream of the shock as well as the spatial structure of the shock as represented by the flow velocity.

EMP presented observations and Monte Carlo simulations of the earth's bow shock derived from satellite measurements made during 1984 September 12, when the interplanetary magnetic field strength was $\sim 100\mu G$ and aligned nearly radially from the sun. Thus, the shock had a quasi-parallel geometry and was similar to the shock subsequently studied by EGBS. Details of the observations are given in EMP. The Monte Carlo techniques were the same as those used in EGBS, and the simulated shock was assumed to be plane parallel. EMP showed observed and simulated particle distributions both upstream and downstream of the shock. Since the bow shock measurements were *in situ*, and there appeared to be some change in the properties of the solar wind in the time interval between the upstream and downstream measurements, EMP modeled two slightly different shocks for the comparisons. Those are indicated in Table 1 by EMPd (downstream measurements: UT=11:26-11:28) and EMPu (upstream measurements: UT=11:32-11:37). In addition, the sonic Mach number of the shock was $\sim 6$ and the Alfvén Mach number $\sim 3$. EMP were able to obtain excellent fits between the observed and simulated particle distributions. We have repeated that experiment using diffusive transport methods.

Since both the hybrid and Monte Carlo simulations are based entirely on direct microphysical kinetic processes, they form shocks of finite thickness related to the dissipation length scales of those processes. They make no explicit distinction between thermal and CR particles, so thermal scattering scales are related fundamentally to those of the CR precursor, or foreshock, that forms in response to propagation of energetic particles upstream of the shock. The foreshock structure is really an integral part of the shock in such models, in fact (Jones & Ellison 1991). Diffusive acceleration simulations, on the other hand, explicitly consider microphysical processes only for the CRs. The diffusive shock precursor is a gasdynamical feature having a characteristic width determined by a balance between upstream CR pressure diffusion (controlled by the mean diffusion coefficient, $\langle \kappa \rangle$) and downstream CR advection (controlled by the velocity $u_1$). This length is effectively determined by the diffusion length, $x_d = \langle \kappa \rangle / u_1$.

Of course, $\langle \kappa \rangle$ is related to the CR particle scattering length as $\langle \kappa \rangle = \frac{1}{3} \langle \lambda v \rangle$, so there is, in fact a connection between the length scales associated with the diffusive transport method and the particle methods once a basis for $\kappa$ is defined. To compare our results with those in EMP and EGBS we need to establish that connection. For hybrid simulations the standard length unit is the ion inertial length, $c/\omega_i = v_A/\Omega_i \propto 1/\sqrt{\rho}$, where $\omega_i$ and $\Omega_i$ are the ion plasma and cyclotron frequencies, respectively. The Monte Carlo simulations depend on a predetermined scattering law. Both EGBS and EMP in their Monte Carlo calculations used a scattering length with a predetermined power-law form, $\lambda = \lambda_o \left(\frac{v}{u_o}\right)^\alpha \left(\frac{n_o}{n}\right)^s$, where $\lambda_o$ becomes the characteristic length needed to define the thickness of the shock, and $v$ is the particle speed. Values for $u_o$ and $n_o$ were taken from the upstream flow ($u_1$ and $n_1$) in EGBS and the downstream flow ($u_2$ and $n_2$) in EMP. Both papers described their spatial structures in terms of $\lambda_o$. That parameter and $\alpha$ were established empirically; from the hybrid simulation in EGBS and from the bow shock in EMP. In EGBS $\alpha = 0.53$, $s = \frac{1}{2}$ and $\lambda_o = 230 c/\omega_i$ were used. EMP used $s = 1$, and obtained best results with $\alpha = 1$. In addition they estimated that $\lambda_o \sim 40 - 400$ km.

Based on these choices and the shock properties given in Table 1 we adopt in our calculations analogous forms for the CR proton scattering length. The associated diffusion coefficient is then determined by equation 2-11. For the EGBS comparison we used

$$\lambda = 5.15 \times 10^9 (\frac{v}{u_1})^{0.53} (\frac{\rho_1}{\rho})^{1/2} \text{ cm}, \tag{2-18}$$

so that $\lambda_o = 5.15 \times 10^4$ km. For the EMP comparison, instead we adopted the form

$$\lambda = 3.13 \times 10^{12} (\frac{\xi}{B}) (\frac{\rho_1}{\rho}) p \text{ cm}, \tag{2-19}$$

where the magnetic field strength, $B$, is expressed in units of $\mu G$. Since this latter form resembles that for Bohm diffusion, $\xi$ is introduced as a scaling constant defining the scattering length in terms of the gyroradius, $r_g$, of the protons; i.e., $\xi = \lambda/r_g$. For our simulations matched to the EMP results we used $B = 100$, $\rho_2 = 4.9$, $u_2 = 115$ km s$^{-1}$, and $\xi = 5$, so that $\lambda_o = 122$ km. Except for physical lengths, which scale directly with $\xi$, the time asymptotic results of our simulations do not depend on the value of $\xi$, however.

An important feature of the simulations reported in EMP and EGBS was the inclusion of a Free Escape Boundary (FEB) that was intended to model the escape of higher energy particles due to the finite extent of the bow shock and/or insufficient scattering far upstream to return particles to the shock. In the hybrid and Monte Carlo calculations the FEB was created by simply removing any upstream facing particle that reached a specified distance, $x_{FEB}$, upstream of the shock. In our diffusive acceleration simulations the FEB is realized by setting $f(p > p_2) = 0$ for the kinetic equation calculations or $P_c = 0$ for two-fluid calculations at all positions upstream of $x_{FEB}$. This procedure provides operational source terms for equations 2-10 and 2-12 at the FEB. The FEB sets an effective limit to the maximum momentum that particles achieve and causes the particle distribution and the dynamical structure of the shock to come fairly quickly to an equilibrium. The Monte Carlo simulations are inherently time independent, but the hybrid and diffusion calculations are not, so this feature facilitates comparisons. On the other hand, the computational mechanisms for realizing the FEB are rather different for the diffusive transport method, so it also provides a challenging test.

## 2.5. Numerical Methods

Some other details about the way our own calculations were conducted may be useful to know. The gas dynamic equations (2.7)-(2.9) are solved using the explicit Piecewise Parabolic Method (PPM) finite-difference hydrodynamic scheme, including the added terms necessary to treat CR pressure. For all simulations the gas adiabatic index is $\gamma_g = 5/3$. Assuming their $\alpha$ parameter for injection to be a constant, Zank, Webb & Donohue 1993 derived an effective equation of state for the thermal plasma. Energy loss to the CRs effectively softens the equation of state within the subshock, or wherever injection is active. Our injection scheme resembles theirs and would be identical in the two-fluid version if $f(p_2)$ were a constant, and there were no Alfvén wave transport included. The latter effect, through wave dissipation, stiffens the effective equation of state within the foreshock region. In both of our numerical schemes these modifications to the equation of state for the plasma are implicitly and self-consistently included by our use of source terms, $S$ or $S_{tf}$ and $L$ in 2-9. Equation 2-10 or 2-12 is solved using the implicit, Crank-Nicholson finite-difference scheme. Details of the numerical schemes can be found in Jones & Kang 1990 and Kang & Jones 1991. Flow was initiated as a simple, plane-symmetric, pure gasdynamic ($f(p > p_2) = 0$ or $P_c = 0$) shock facing to the right. We have chosen a reference frame approximately at rest with respect to the shock. Since we start the simulation with no CRs, and the total shock compression increases in response to building CR pressure, the shock speed relative to the upstream gas will usually decrease with time as the compression approaches the steady state. Thus the initial hydrodynamic shock speed should be slightly larger than the desired final shock speed, $u_1$. The exact initial value can be determined only numerically by iteration, but the correction factor is less than 10 %, so that it can be ignored in the discussion of results. Except for CRs beyond the FEB (set to zero), all quantities are treated as continuous on the grid boundaries. We assumed the thermal proton pressure was dominant over the thermal pressures from all other ions and the electrons in the downstream region, while the thermal protons and electrons contribute equal pressures in the upstream region. This approximately accounts for the observation that the electrons are heated much less strongly than ions through the shocks. To simplify our calculation and keep the computational costs down we included only protons in the CR population, as discussed in §2.2. For the kinetic equation simulations we used 128 momentum cells logarithmically spaced from $p_o = 10^{-4}$ to $p_3 = 10^{-1}$. This resolution has proved to be sufficiently fine to produce converged solutions with our methods (Kang & Jones 1991).

It is preferable in the code to work with physical variables that are normalized by "natural units", so that the values in the computer are neither very large nor very small. Thus, the shock structure information we display is most easily presented in those normalized units. The normalization constants for the EMP tests are: $n_o = 1$ cm$^{-3}$, $t_o = 60$ s, $u_o = 575$ km s$^{-1}$, $r_o = 3.45 \times 10^4$ km, $P_o = 5.52 \times 10^{-9}$ dyne cm$^{-2}$, and $\kappa_o = u_o r_o = 1.98 \times 10^{17}$ cm$^2$ s$^{-1}$. $n_o$ and $t_o$ remain the same in the EGBS test. Otherwise the normalization constants used for the EGBS simulations are: $u_o = 378$ km s$^{-1}$, $r_o = 2.27 \times 10^4$ km, $P_o = 2.39 \times 10^{-9}$ dyne cm$^{-2}$, and $\kappa_o = u_o r_o = 8.57 \times 10^{16}$ cm$^2$ s$^{-1}$.

As mentioned earlier, the CR injection model we use with the kinetic equation depends on keeping the numerical shock thickness comparable to the "real" shock thickness, which we take as a few times $r_g(p_{th})$. To maintain that constraint we have used a spatial zone size similar to the gyroradii, $r_g$, of particles with $p_{th}$ (based on the postshock temperature estimated from the initial hydrodynamic shock). This reflects the fact that numerical shocks spread over 2-3 zones in the PPM code. As expressed in normalized units, $\Delta x = 7.5 \times 10^{-3}$ for the EMP models and $\Delta x = 0.1$ for the EGBS model. One can easily understand that the numerically realized injection rate will vary with the numerical parameters $\Delta x$ and $c_1$, or more directly, the ratio $\lambda(p_1)/\Delta x$. The assumed scattering laws (equations 2-18 and 2-19) also influence this, because they determine how fast particles in the injection pool can gain energy. The ratios are $\lambda(p_1)/\Delta x \approx (10\rho^{-1})$ for EMP and $\lambda(p_1)/\Delta x \approx 30\rho^{-0.5}$ for EGBS, where $\rho = 1$ for the upstream gas. Thus, since we have established elsewhere that convergence with these methods requires between 10 - 20 zones within the length $x_d$ Kang & Jones 1991, Frank, Jones & Ryu 1995, the diffusive transport of the particles with $p_1$ may be only marginally converged for the EMP case. But, since the equations themselves are only qualitatively valid near this momentum range, our aim is not to follow the diffusion and acceleration of these particles accurately. Rather, it is to model approximately the injection rate at $p_2$ by adjusting the free parameter $c_1$. In test runs with zone sizes half the values given above, slightly larger values of $c_1$ were required to obtain the similar compression through the shock and CR energy density. In this regard, a numerical hydrodynamic scheme that utilizes an artificial viscosity that produces a shock form closer to a real viscous shock structure might be better suited for this type of calculation. That is provided one can find an artificial viscosity that has a physical basis in collisionless shocks. The grid in the EMP simulations extended over an $x$ interval [-94.2, 110.6], with the shock initially located at $x = 85$, while the EGBS grid was on an interval [0,15.36], with the shock positioned at $x = 13.45$.

## 3. Test Results

We reiterate that our goals in this study are to evaluate quasi-parallel shocks simulated with the diffusive model for energetic particle transport by comparing such simulations to models computed by independent, particle methods and to a real, observed collisionless shock. Since it is more complete, we will start with the kinetic equation version of the model. This allows us the opportunity to compare the particle distribution functions, as well as the dynamical structures of the shocks. In each test we have assumed the basic shock parameters listed in the first three columns of Table 1 and the subsidiary information about the shocks from EGBS and EMP as discussed in the previous section. The aim in each test was to match as closely as possible the reported forms of the proton distribution function and the total compression through the shock structure. In doing this we allow two free variables. The first is the fiducial momentum, $p_1 = c_1 p_{th}$, which controls the rate for particle injection into the high energy, CR population. Second, the displacement of the FEB from the gas subshock is varied from the values in EGBS and EMP in order to take account of the difference in the model realization of the FEB from theirs. All other properties of the simulations are fixed as outlined earlier. In the particle simulations, on the other hand, the location of the FEB is the only free parameter. To illustrate the significance of Alfvén wave transport we did simulate each shock both including and excluding those terms, but we believe the models including Alfvén wave transport are the more meaningful ones.

As a follow-up we have then recomputed the evolution of these same shocks using a two-fluid version of the diffusive transport model. Those simulations were based on closure parameters estimated from the kinetic equation tests. Except as required by the practical differences in these two version of the diffusive transport model the two-fluid simulations we present are exactly analogous to the kinetic equation simulations.

### 3.1. Kinetic Equation Results

Particle distribution information in EGBS and EMP was presented in terms of the omnidirectional particle flux measured in the shock frame. Therefore, we will do the same in our comparisons. The distribution function in the diffusion approximation is nearly isotropic in the local scattering center frame. Thus the omnidirectional particle flux per unit solid angle per unit kinetic energy in the shock frame can be calculated from the computed distribution, $f(p)$, according to

$$F(E') = \frac{1}{2} \int_{-1}^{1} d(cos\theta) p^2 f(p) \frac{dp}{dp'}, \tag{3-1}$$

where the particle momenta $p$ and $p'$ are defined in the local scattering frame and in the shock frame, respectively, and $E' = \frac{1}{2} m_p c^2 p'^2$ is the particle kinetic energy in the shock frame. For nonrelativistic flows the momenta in the two reference frames are related by $p'^2 = p^2 + (\frac{u}{c})^2 + 2\vec{p} \cdot (\frac{\vec{u}}{c})$ where $\vec{u}$ is the bulk flow velocity in the shock frame. Except for some modest corrections at the lowest energies there is little difference in correcting the flow speed in equation 3-1 to include $\vec{u}_w$. But, since it is much simpler to ignore that correction, we will.

#### 3.1.1. Comparison with EGBS particle methods

Fig. 2 and Fig. 3 illustrate a comparison between shock properties found in our diffusive transport simulations and those reported in EGBS. Recall that our simulations are time dependent and that we began with a pure gasdynamic flow, so that the particle distribution was everywhere Maxwellian and the shock was a simple fluid discontinuity. By about $t = 80$ minutes our shocks have reached an approximate steady state in terms of the dynamical properties and the proton distribution (see Fig. 1 for an indication of how $f(p)$ evolves). For the diffusion coefficient used in this test the characteristic diffusion time, $t_d = \kappa(p)/u_1^2 < 10$ minutes, for relevant momenta. As we will see in the discussion of two-fluid models the location of the FEB is roughly a diffusion length, $x_d$, upstream of the subshock. Thus, both energy and particles will begin to escape in significant amounts on a timescale of a few minutes and it is, therefore, reasonable to expect equilibrium to develop on a similar time. The data from our simulations shown in Fig. 2 and Fig. 3 are for $t = 80$ minutes. The solid curves represent results for a kinetic equation simulation including Alfvén wave transport while the dashed curves represent a simulation omitting those terms. The dot-dashed curve in Fig. 3 represents results from a two-fluid test to be discussed in §3.2. Fig. 2 provides a comparison between the omnidirectional fluxes computed by us at positions listed in Table 1 and those presented for both particle methods in EGBS. The left panel illustrates conditions downstream of the subshock (EGBSd), while the right panel shows the particle distribution just upstream of the subshock (EGBSu). The particle simulation data from EGBS are shown as filled circles (Monte Carlo) and open circles (hybrid), and are representative values read from Figures 1 and 2 in EGBS. For consistency with the stated model parameters we must assume that the EGBS fluxes were actually normalized by the shock velocity, although it was not explicitly mentioned in that paper. Fig. 3 shows the shock structure in terms of gas density, pressure and velocity along with the CR pressure. The velocity structure from the EGBS hybrid simulation (from figure 3 in EGBS) is again shown as open circles. We chose model parameters to keep the total compression through the structure as close as we could to the value 4.66 fixed by EGBS for both types of particle simulation. Our simulations shown achieve this within about 1% as illustrated in Figure 3. Flow structures far downstream of the subshock are dependent on initial conditions, since these are time-dependent calculations, and should be ignored in evaluating the results.

For our model shown in Fig. 2 and Fig. 3 that included Alfvén wave transport the parameter $c_1 = 1.87$, so that the particle injection pool was bounded below by momenta 1.87 times the peak in the postshock thermal distribution. The simulation done without Alfvén wave transport required a slightly higher value, $c_1 = 1.95$. In effect a larger injection pool was needed for the case with Alfvén wave transport to compensate for the slower acceleration rate resulting from a reduced advection rate for CR and a reduced subshock compression (cf. Jones 1993). In either case, as we will see in the discussion of the analogous two-fluid model, the flux of particles being injected to the CR population is $\sim 2 - 3\%$ of the total passing through the shock.

The upstream displacement of the FEB in our models is listed in Table 1. For best results, especially for matching the distribution functions above $\sim 50$ kev, we require an $x_{FEB}$ about 20% greater than the value used in EGBS for their Monte Carlo simulation. Smaller values of $x_{FEB}$ increase the rate of particle escape, so that equilibrium is achieved more quickly and with fewer high energy particles and smaller total $P_c$. This discrepancy in the location of the FEB probably reflects a subtle but significant difference in the physical meanings of the FEB conditions we apply. Our methods, based on an assumed *isotropic* distribution function, force the entire CR population to zero at the FEB, and so perhaps reduce a little too abruptly the full population of particles just on the shock-ward side of the FEB. Thus, to mimic closely the behavior of the Monte Carlo FEB, which eliminates only the outward bound CR, we need to place our formal boundary slightly farther from the shock.

With these parameters our dynamical and distribution results are generally in very good agreement with those of both particle methods in this test. As shown in the velocity profile of Fig. 3 our shock structure comes quite close to that of the hybrid simulation of EGBS. The Monte Carlo results (not shown here, but illustrated in EGBS) exhibit a somewhat broader shock profile. The fact that the total shock compression is $\approx 4.7$, rather than 4 or less, as expected for a gas with an adiabatic index equal to 5/3, is due entirely to the FEB. Because of the energy loss through the FEB these shocks act like radiative shocks. In terms of the particle distributions, we get slightly better agreement with the Monte Carlo simulation downstream of the subshock. The gas temperature is slightly hotter than the hybrid simulation result. On the other hand, our distribution of suprathermal particles (the lowest energy CRs) is closer to that of the hybrid calculation. Upstream, the hybrid simulation particle fluxes shown represent results for a position $0.4\lambda_o$ from the subshock. To get good agreement we found it necessary to choose a location $3\lambda_o$ from the subshock, just as EGBS did for their Monte Carlo simulation. Our flux distribution at $0.4\lambda_o$, like the EGBS Monte Carlo result, has a much broader form. Thus, in this respect we are in closer accord with the Monte Carlo model. Perhaps this reflects the fact that both the diffusion model and the Monte Carlo model depend on simple, approximate, *a priori* scattering relations for the particles while the hybrid simulation computes a locally self-consistent wave field and follows particles exactly in that field. On the other hand, it is conceivable that by including a self-consistent treatment of the growth of the Alfvén wave field as discussed in Jones 1993, we would obtain results closer to the hybrid calculations, since the scattering strength would be locally determined. We have not attempted to do this, however.

### 3.1.2. Comparison with EMP Bow shock measurements

The previous test demonstrates general consistency between the properties of shocks studied through the kinetic equation version of the diffusive transport model for CRs and the properties of shocks computed by hybrid plasma and Monte Carlo models. Ideally, however, we need to compare models with real shocks. The best available opportunity to apply these models to a real shock is through the bow shock results reported in EMP. Our procedure in this case was the same as for the previous one, and results are illustrated in Fig. 4 and Fig. 5. As before, the curves illustrate properties of our simulations (same symbols as for Figs. 2 and 3). The filled circles in Fig. 4 are representative bow shock flux values read from figure 6 in EMP. That EMP figure also presents comparative Monte Carlo simulation results, but we will not discuss those here, except to say that they are very similar to our own. As before we include results both for simulations including Alfvén wave transport and omitting it. Our results at $t = 8$ minutes will be compared with the EMP observations. For our simulated bow shocks the distribution function and the flow structure become steady around $t = 2$ minutes. Under these circumstances this is the acceleration time scale to reach the maximum particle energy ($\sim$ 100 keV), which in turn is determined by the distance of the FEB from the shock front and $\kappa(p)$. Without a FEB the acceleration continues to higher energy and the CR pressure increases for a much longer time. As with the EGBS test, in order to match the particle flux above $E > 50$ KeV, we had to use a larger value of $x_{FEB}$ (40 % greater in this case) than that which worked best for their Monte Carlo simulations.

Recall in this case that the EMP observations seemed to indicate some change in the flow conditions between the upstream and downstream observations. EMP argued, and we would agree, that the time interval between the two observations was long enough for the bow shock to reestablish an equilibrium. Therefore, we modeled two separate shocks (Table 1). The structure in Fig. 5 comes from our simulations of the EMPd shock, although the properties of the EMPu shock are quite similar. The omnidirectional proton flux in the shock frame was calculated at the downstream position for the model EMPd shock and at the upstream position for the model EMPu shock at positions listed in Table 1. EMP made their upstream comparison at $40\lambda_o$ in the Monte Carlo simulation.

Our best results were obtained using $c_1 = 1.5$ with Alfvén transport included and $c_1 = 1.7$ when it was excluded. The difference indicates again a requirement for a larger injection pool of particles to compensate for the acceleration inhibitions imposed by Alfvén wave transport. In our comparison two-fluid calculation we will see that the flux of injected particles $\sim 10\%$ of the total in this case. The equilibrium CR population represents $\sim 2.5 - 3\%$ of the total downstream density, consistent with what was observed by EMP. Compared to the EGBS case the lower bound of the injection pool ($c_1$) is about 20% smaller, so a greater portion of the particles in the shock make it into the CR population. This difference results from the different forms of scattering length adopted in the two models (see Equation 2-18 and 2-19). The thickness and structure of the physical subshocks in EGBS and EMP would presumably be different due to the different scattering laws. Consequently, the effective momentum above which particles should "leak" in our injection model should be different as well. Recall that to minimize the number of free parameters we fixed the thickness of the numerical shock at 2-3 times the gyroradius for $p_{th}$.

It is clear from Fig. 4 that the diffusive transport model does a very adequate job of simulating the particle distribution in this shock. The only apparent discrepancy is a modest deficiency of suprathermal particles upstream, very much like that noted in EMP for their Monte Carlo simulation. EMP did not attempt to establish the structure of the measured bow shock in any detail, so we cannot compare our Fig. 5 with that shock. We include it for completeness and for comparison with two-fluid results described in the next section.

### 3.2. Two-Fluid Results

In order to examine the ability of the two-fluid model of diffusive CR acceleration to follow accurately the dynamics of quasi-parallel shocks, we have repeated the calculation of shock structure for both the EGBS and EMP cases with the two-fluid model, using information from the kinetic equation calculations to define the necessary closure parameters. We argued in the introduction that the primary challenge to useful application of the two-fluid model was *a priori* determination or estimation of these closure parameters. For the shocks under study here we have a means to do this properly, so we can test this point of view. We can also examine in these instances how sensitive the results are to the selection of closure parameters.

For all our two-fluid simulations here the CR adiabatic index was assumed to be $\gamma_c = 5/3$, since the particles remain non-relativistic throughout each calculation. We present only the cases with Alfvén wave transport, since nothing new is added by comparing results here that do not include Alfvén wave transport. The principal additional quantities that need to be defined for these comparisons are the diffusion coefficient, $\langle \kappa \rangle$, the location of the FEB and a suitable two-fluid version of the injection model. We have already discussed the injection model in §2.1.

Ideally the mean diffusion coefficient we use here should be determined from equation 2-13. However, the integrands in 2-13 are actually difficult to use in practice, because the numerical spatial derivatives are not meaningful through the shock. Thus, we actually applied the common, approximate expression,

$$\langle \kappa \rangle = \frac{\int_{p_2}^{p_3} f\kappa[\sqrt{p^2+1}-1]p^2 dp}{\int_{p_2}^{p_3} f[\sqrt{p^2+1}-1]p^2 dp}, \qquad (3\text{-}2)$$

which assumes one can separate the $x$ and $p$ dependence in $f$. Although that is not really valid, we find from our results that it is an adequate approximation. For both the EMP and EGBS simulations equation 3-2 applied to the kinetic equation results led empirically to a $\langle \kappa \rangle$ that could be well represented by the expression

$$\langle \kappa \rangle = \{\kappa_a + \kappa_b \left[1 - \exp(-\frac{t}{t_o})\right]\}\{\rho_1/\rho\}^s, \qquad (3\text{-}3)$$

given in the normalized units defined in §2. The index, $s$, is $\frac{1}{2}$ for EGBS and 1 for EMP, as discussed in §2. For the EMP test $\kappa_a = 0.1$, $\kappa_b = 0.23$ and $t_o = 8$, while for the EGBS test $\kappa_a = 3.4$, $\kappa_b = 5.95$ and $t_o = 32$. Thus, $\langle \kappa \rangle$ increases rapidly from an initially small value, but then approaches a steady value. This behavior is easy to understand both qualitatively and semi-quantitatively. First, the initial small value reflects the facts that for both cases $\kappa(p)$ is a strongly increasing function of $p$ and that at the start $\langle \kappa \rangle$ should be strongly weighted towards $\kappa(p_2)$. Then $\langle \kappa \rangle$ increases as particles are accelerated to higher energies, so that $f(p > p_2)$ develops a significant population. On the timescale that particles are accelerated to momenta such that $\kappa(p_m)/u_1 \approx x_{FEB}$ the distribution function begins to approach an equilibrium (see Fig. 1), and $\langle \kappa \rangle$ also approaches an asymptotic value. Thus, as observed, the timescale for $\langle \kappa \rangle$ to evolve is the same as for the flow to approach equilibrium. The final, steady value of $\langle \kappa \rangle$ is such that $x_d = \langle \kappa \rangle/u_1 \approx x_{FEB}$, as one might expect. Indeed, we found that as long as the ratio $r_d = x_{FEB}/x_d$ was fixed, the time asymptotic postshock state was independent of the actual value of either $x_{FEB}$ or $x_d$ for the two-fluid models. On the other hand the solutions were quite sensitive to $r_d$, since this controls the relative ease with which CR energy escapes the system. The size of the precursor and timescale required for the shock to reach a dynamical equilibrium were directly proportional to $x_d$, however. The results we show in Fig. 3 (EGBS test) and in Fig. 5 (EMP test) used the same values of $x_{FEB}$ that are given in Table 1 and that were used in the kinetic equation simulations. These correspond to $r_d = 1.28$ in the EGBS case and $r_d = 1.15$ in the EMP case.

The two-fluid results are illustrated by the dot-dashed curves in Fig. 3 and Fig. 5 at the same times as the kinetic equation results shown. It is clear that the agreement between the two methods is excellent. The individual postshock pressures, $P_g$ and $P_c$, agree to better than 2% of the total momentum flux, $P_g + P_c + \rho u^2$, which, within our experience, is consistent with the general accuracy limits of our methods.

Up to now all comparisons have been devoted to evaluation of the time asymptotic properties of the various shocks, because that is all that is available from EGBS and EMP. However, since both our diffusive transport methods are time dependent, and since for many astrophysical applications one cannot expect steady state conditions to prevail, it is important to evaluate the agreement between the time evolution of shocks computed with the kinetic equation and two-fluid models. Fig. 6 provides that comparison for the EGBS shock structure. Here we show for both diffusive transport methods the computed shock structures at $t = 0$, 1, 4, and 8 minutes. The same line types are used as before. The initial shock was at $x = 85$ and it drifts slowly to the left over time. The evolution of the two shock structures are in very good agreement. The slight difference in the subshock position at later times comes from the fact that CR acceleration begins slightly faster in the two-fluid model, so that the total compression rises a little sooner in that case. This is visible in the postshock density profiles, which approximately preserve the early histories of the shock evolution. The small amplitude postshock fluctuations in gasdynamic variables are numerical in origin and result from small fluctuations in the transfer of energy between CRs and gas.

Experiments we have conducted show that we could obtain fairly similar two-fluid evolution with simple "intuitive" models for the closure parameters. For instance, one can derive a simple evolutionary model of the mean diffusion coefficient, $\langle \kappa \rangle$, by using the stated form for $\kappa(p)$ along with simple power-law estimates of the form of $f(p < p_{max}(t))$ based on the strength of the subshock and the maximum expected momentum, $p_{max}(t)$, derived from standard analytic estimates of the acceleration time (e.g., Lagage & Cesarsky 1983). The existence of the FEB would be considered by imposing a maximum value for $p_{max}(t)$ from the constraint $\kappa(p_{max})/u_1 \leq x_{FEB}$. This leads to a $\langle \kappa \rangle$ that changes over time with a form qualitatively similar to equation 3-3. Although the evolutionary properties of such simple models differ in fine detail from that shown in Fig. 6, they certainly would be adequate for making the comparisons we have conducted here and for addressing such issues as the expected structures and acceleration efficiency of these shocks.

## 4. Conclusion

In this paper, we have carried out several nonlinear numerical simulations of diffusive acceleration of cosmic-ray particles at quasi-parallel shocks, and made comparisons with particle simulations and bow shock measurements. We computed the time evolution of these shocks using both the momentum dependent, kinetic-equation or diffusion-advection version of diffusive transport theory and the simpler two-fluid version. Both versions account for dynamical feedback between the high energy particles and the thermal plasma. To account for injection of thermal protons into the population of high energy particles in shocks we have adopted a simple model for use in the kinetic equation in which a population of diffusive particles is introduced at intermediate momenta between the plasma and cosmic-ray particles in order to model "thermal leakage" from the high energy tail of the thermal, Maxwellian distribution. The properties of this "injection pool" are controlled in our model by one free parameter that is adjusted in order to identify particles whose mean scattering lengths exceed by a factor of several the numerical shock thickness. We used results from this model to simulate the same processes in two-fluid versions of each computation.

Previous work of Ellison *et. al.* (1993) had carried out a comparison of calculations designed to demonstrate that hybrid plasma simulation and Monte Carlo simulation give comparable results. We have extended that comparison by simulating the same shock using both versions of diffusive shock acceleration theory. By adjusting the free parameter in our injection model over a very restricted range, we were able with the kinetic equation method to generate an omnidirectional particle flux and a flow velocity through the shock (or equivalently a density profile through the shock) that are consistent with theirs.

We have also applied our methods to a comparison with observations of the earth's bow shock when the magnetic field had a quasi-parallel geometry. We showed that with our simple injection model the diffusive transport model can reproduce the proton flux spectrum observed in this shock and presented by Ellison *et. al* (1990). Again the range of variation needed in the free parameter for injection is very small. Exploitation of more sophisticated techniques that incorporate explicit self-consistent scattering wave fields may enable us to further refine this model so that the value of the free parameter is even better defined.

Using the properties of the kinetic equation diffusive transport formulation for guidance in determining the closure parameters for a two-fluid version of the same calculation we also showed that the kinetic equation and two-fluid method give consistent solutions to these shock flows.

These results demonstrate that all three momentum dependent techniques obtain similar shock and particle distribution properties and that the shock structures and particle acceleration properties obtained through the two-fluid approach are also consistent with the others. This is despite some fairly different assumptions made by each of the computational methods. The diffusive transport theory, for instance, assumes that the cosmic rays are a distinct, massless population whose momentum distribution is isotropic in the center-of-momentum frame of a posited field of scattering Alfvén waves, and that the response of high energy particles to those waves can be treated as spatial diffusion. The two-fluid model goes one step further and explicitly follows only the energy of the cosmic-rays, dealing implicitly with the form of the particle distribution through *a priori* closure parameters (which may, however, vary in space and time). Both hybrid and Monte Carlo methods consider the particles to belong to a single population and follow them explicitly (at least for protons). On the other hand hybrid methods attempt to solve the equations of motion and the electromagnetic field equations self-consistently, while Monte Carlo methods assume an *a priori* scattering environment for the particles, which are followed through discrete encounters. That all of these methods seem to agree in the basic outcome is both remarkable and very advantageous. It means that the essential physics of particle acceleration is present in each of them, and that they are all practical and complementary tools for understanding the whole of the physics of collisionless shocks and their roles in astrophysical environments.

One of our original motivations for this set of experiments was to help resolve controversies that persist over the use of two-fluid methods in studies of diffusive acceleration, as outlined above. We believe that our results add credibility to their use as a means to explore dynamical evolution of CR modified shocks. In particular these results should help answer questions about the applicability of two-fluid methods when it is not proper to assume a momentum independent diffusion process, or about the ability of the methods to deal with the consequences of escaping particles, even though particle number is not directly incorporated into the theory. Much of the past basis for those concerns was derived from properties of the time independent version of the theory as it applies to an infinite, one dimensional space. In that setting the issues identified take on a different meaning that comes from the fact that the assumption of time independence becomes unrealistic itself. Although the present simulations approach a time independent state, that is possible because of the existence of the Free Escape Boundary that removes both particles and energy from the computational space. It is clear that two-fluid methods deal with that situation adequately. We must hasten to emphasize, as always, however, that the accuracy of two-fluid methods depends on our ability to predict appropriate forms for the diffusion processes and their effect on the relative portions of relativistic and nonrelativistic particles, so that we can predict the behavior of the cosmic-ray adiabatic index. Finally, we mention that in related work, we have begun to address some of the same issues as they apply to cosmic-ray modified shocks with oblique magnetic field geometries. Some preliminary results are presented in Frank, Jones & Ryu 1995.

We would like to thank Peter Duffy and Larry Rudnick for helpful comments on the manuscript. HK was supported at Pusan National University by the Korean Research Foundation through the Brain Pool Program. This work was supported in part at the University of Minnesota by the NSF through grants AST-9100486 and AST-9318959, by NASA through grant NAGW-2548 and by the University of Minnesota Supercomputer Institute.

― 26 ―Table 1. Shock Model Parameters[a]

| Model | $u_1$ (km s$^{-1}$) | $n_1$ (cm$^{-3}$) | $T_1$ (K) | $x_{FEB}$[b] | $D_{obs}$[b] |
|---|---|---|---|---|---|
| EGBSd | 378 | 1.04 | $1.24 \times 10^5$ | $5.3\lambda_o$ | $-0.44\lambda_o$ |
| EGBSu | 378 | 1.04 | $1.24 \times 10^5$ | $5.3\lambda_o$ | $3.0\lambda_o$ |
| EMPd | 575 | 1. | $2 \times 10^5$ | $144\lambda_o$ | $-4.2\lambda_o$ |
| EMPu | 575 | 1.7 | $2 \times 10^5$ | $144\lambda_o$ | $54\lambda_o$ |

[a]Our sign convention for the spatial coordinate is reversed from EMP and EGBS. Thus, our velocities and spatial displacements have the opposite sign to theirs.

[b]$\lambda_o = 5.14 \times 10^4$ km for EGBS and $\lambda_o = 122$ km for EMP models.

# REFERENCES

Achterberg, A., Blandford, R., & Periwal V., 1984 A&A, 132, 97

Axford, W. I., 1981, Proc. 17th Int. Cosmic Ray Conf. (Paris), 12, 155

Axford,W. I.,Leer, E.,& Skadron, G. 1977,*Proc. 15th Internat. Cosmic-Ray Conf.* (Plovdiv), 11, 132

Baring, M. G., & Kirk, J. G., 1991, A&A, 241,329

Bell, A. R., 1978, MNRAS, 182, 147

Bell, A. R., 1987, MNRAS, 215, 615

Berezhko, E. G., & Krymskii, G. F., 1988, Soviet Phys. Usp., 31, 27

Blandford, R. D., & Ostriker, J. P., 1978, ApJ, 221, L29

Blandford, R. D., & Eichler, D., 1987, Phys. Rept., 154, 1

Donohue, D. J., Zank, G. P., & Webb, G. M., 1994, ApJ, 424, 263.

Dorfi, E. 1984, Adv. Space Res., 4, 205

Dorfi, E. A., 1990, A&A, 234, 419

Dorfi, E. A., 1991, A&A, 251, 597

Drury, L. O'C., 1983, Rept. Prog. Phys., 46, 973

Drury, L. O'C., & Falle, S. A. E. G., 1986, MNRAS, 223, 353

Drury, L. O'C., Markiewicz, W. J., & Völk, H. J., 1989, A&A, 225, 179

Drury, L. O'C., & Völk, H. J., 1981, ApJ, 248, 344

Duffy, P. 1992, A&A, 262, 281

Duffy, P., Drury, L. O'C., & V"olk, H. 1994, A&A, 291,613

Eichler, D. 1979, ApJ, 229, 419

Ellison, D. C., Möbius, E., & Paschmann, G. 1990, ApJ, 352, 376 (EMP)

Ellison, D. C., Giacalone, J, Burgress D., & Schwartz, S. J. 1993, JGR, 98, 21085 (EGBS)

Ellison, D. C., & Eichler D. 1984, ApJ, 286, 691

Falle, S. A. E. G., & Giddings, J. R., 1987, MNRAS, 225, 399

Frank, A., Jones, T. W. & Ryu, D. 1995, ApJ, (March 10)

Giacalone, J., Burgess, D., Schwartz, S. J. & Ellison, D. C. 1993, ApJ, 402, 550

Jones, F. C, & Ellison, D. C., 1991, Space Science Reviews, 58, 259

Jones, T. W., 1993, ApJ, 413, 619

Jones, T. W., 1994, ApJS, 90, 969

Jones, T. W., & Kang, H., 1990, ApJ, 363, 499

Jones, T. W., & Kang, H., 1992, ApJ, 396, 575

Jones, T. W., & Kang, H., 1993, ApJ, 402, 560

Jones, T. W., Kang, H., & Tregillis, I. L., 1994, ApJ, 432, 194

Kang, H., 1993, J. Korean Astro. Soc., 26, 1

Kang, H., & Drury, L. O'C., 1992, ApJ, 399, 182

Kang, H., & Jones, T. W., 1990, ApJ, 353, 149

Kang, H., & Jones, T. W., 1991, MNRAS, 249, 439

Kennel, C. F., Edmiston, J. P. & Hada, T., 1985, in *Collisionless Shocks in the Heliosphere: A Tutorial Review*, ed. R. G. Stone & B. T. Tsurutani (Amer. Geophys. Union: Washington, D. C., p 1

Lagage, P. O. & Cesarsky, C. J., 1983, A&A, 125, 249

Markiewicz, W. J., Drury, L. O'C., & Völk, H. J., 1990, A&A, 236, 487

Parker, E. N., 1965, Planet. Sp. Sci., 13, 9

Quest, K. B., 1988, Proc. Sixth Int. Solar Wind Conf., T. O'Neil and D. Brook, NCAR/TN-360+Proc, 503

Ryu, D., Kang, H. & Jones, T. W., 1993, ApJ, 405, 199

Scholar, M. 1990, Geophys Res Lett, 17, 11

Skilling, J., 1975, MNRAS, 172, 557

Völk, H. J., Drury, L. O'C., & McKenzie, J. F., 1984, A&A, 130, 19

Zachary, A. L., Cohen, B. I., Max, C. E. & Arons, J., 1989, JGR, 94, 2443

Zank, G. P., Webb, G. M. & Donohue, D. J. 1993, ApJ, 406, 67

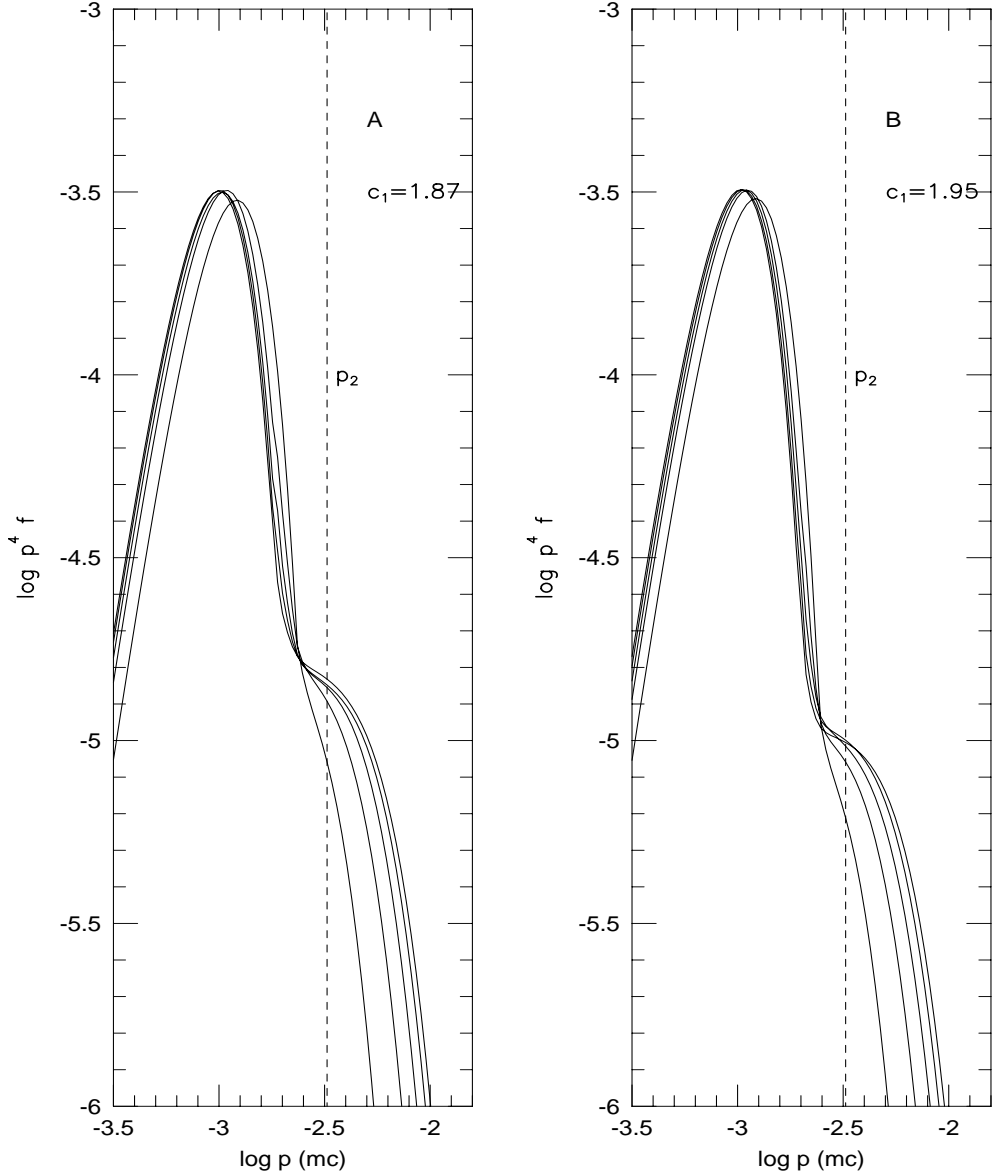

Fig. 1.— Distribution function $g = fp^4$ downstream from the shock for EGBS test runs at $t = 10, 20, 30, 40$ and $50$ minutes. The dashed line marks the momentum $p_2$ at 5 keV above which the particles are considered to be cosmic-rays in the diffusive transport theory. The left panel shows the case with Alfvén wave transport terms included and $c_1 = 1.87$, while the right panel shows the case without such terms and $c_1 = 1.95$.

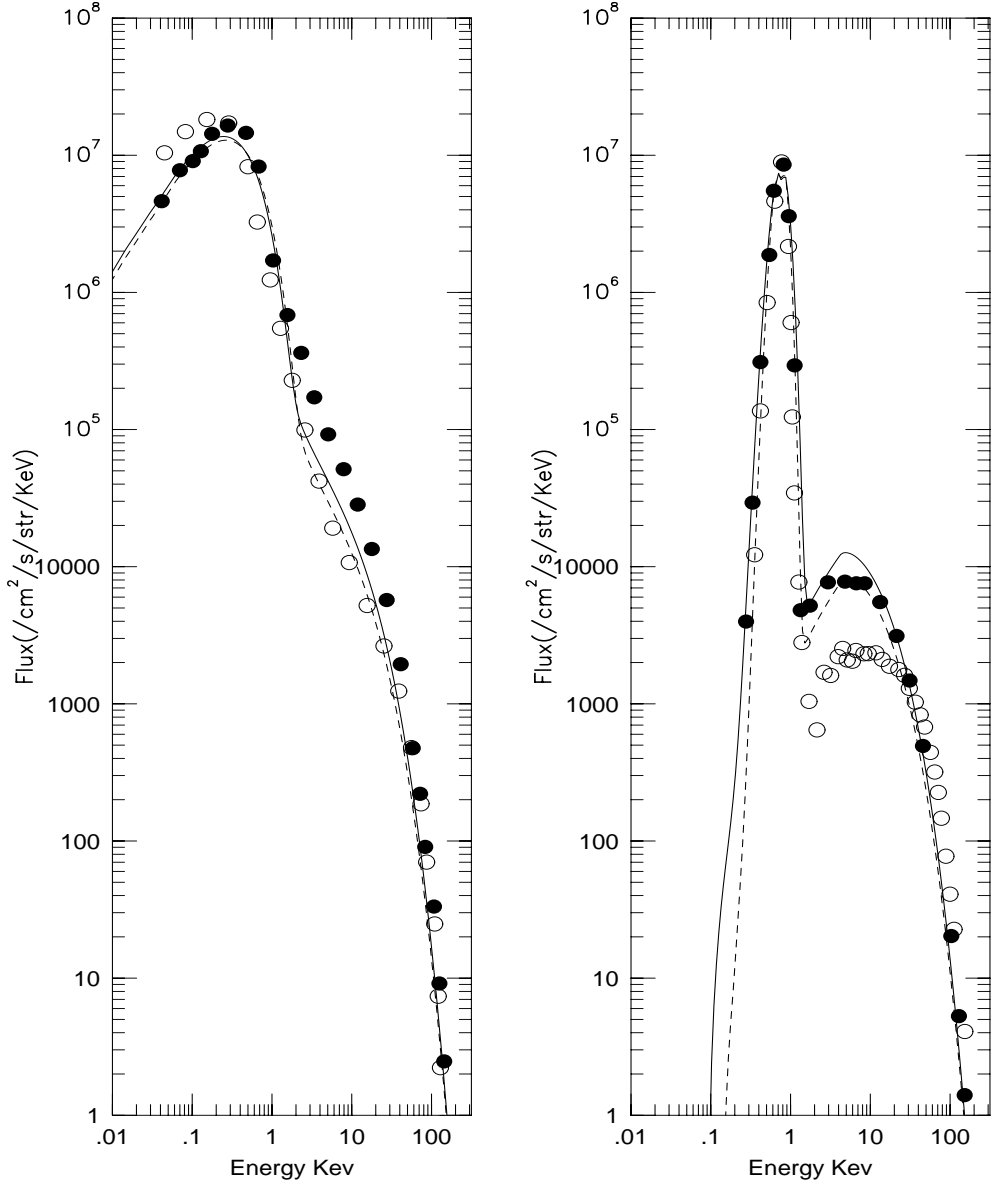

Fig. 2.— Omnidirectional fluxes of protons computed a) downstream and b) upstream of the model EGBS shock according "particle methods" as reported by EGBS and by diffusive shock theory as computed by us. Filled circles are sample Monte Carlo data points, while open circles are from hybrid plasma simulations of EGBS. Our numerical results are shown at $t = 80$ min. The solid lines are for the case with Alfvén wave transport terms included and $c_1 = 1.87$. The dashed lines are for the case without Alfvén wave transport terms and $c_1 = 1.95$.

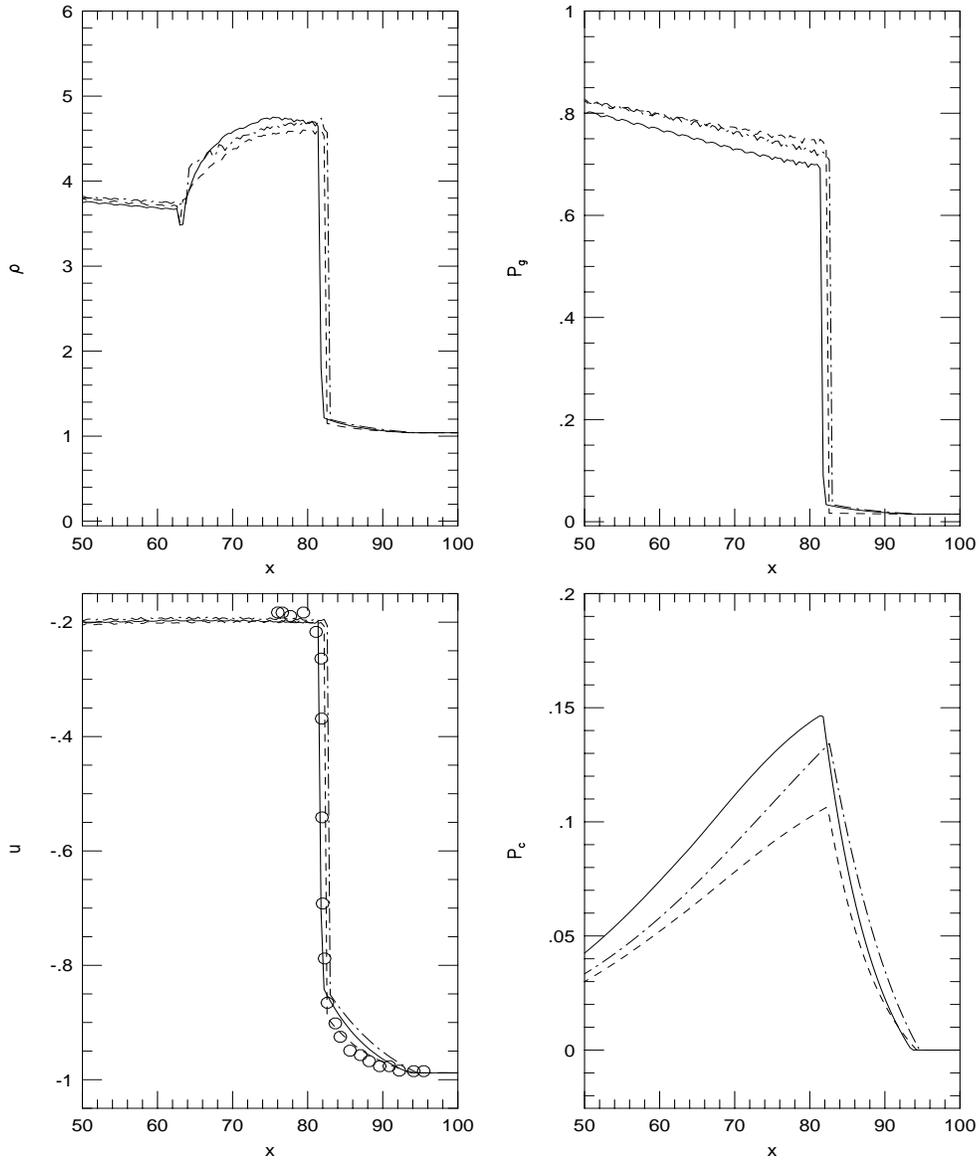

Fig. 3.— The shock flow structure of the model EGBS shock at $t = 80$ minutes in our simulations and from hybrid simulations as reported in EGBS. The line types for the kinetic equation solutions are the same as Fig. 2. The two-fluid solution is shown by the dot-dashed line. The velocity structure from the hybrid method is also shown by open circles as in Fig. 2.

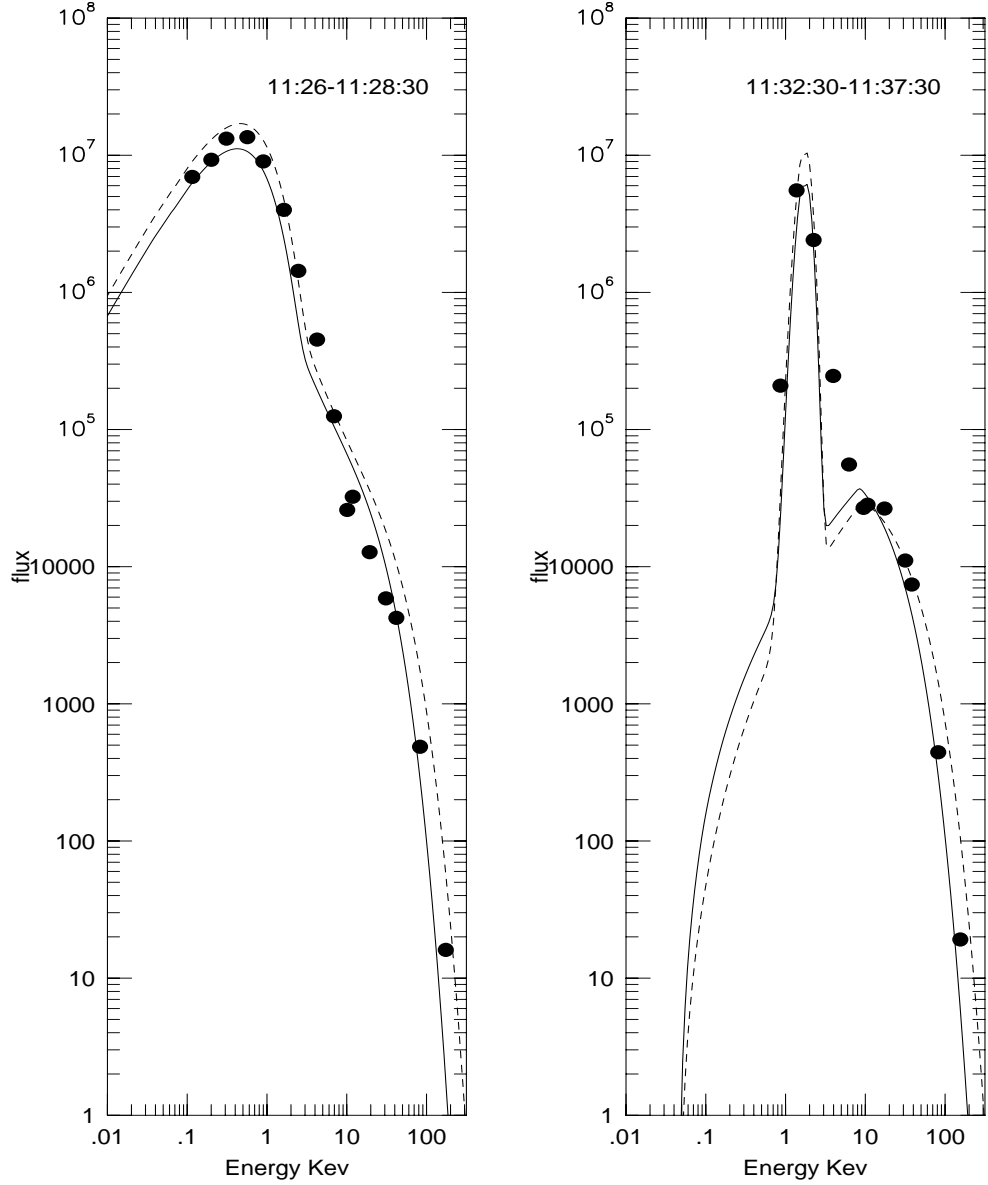

Fig. 4.— Omnidirectional flux of the protons observed and computed a) downstream of the model EMPd shock (observation from UT=11:26) b) upstream of the model EMPu shock (observation from UT=11:32). Filled circles are samples of observed data. The numerical results from our work are shown at $t = 8$ min. The solid lines are for the case with Alfvén wave transport terms included and $c_1 = 1.5$. The dashed lines are for the case without Alfvén wave transport terms and $c_1 = 1.7$. The dot-dashed curves represent the two-fluid diffusive transport solution.

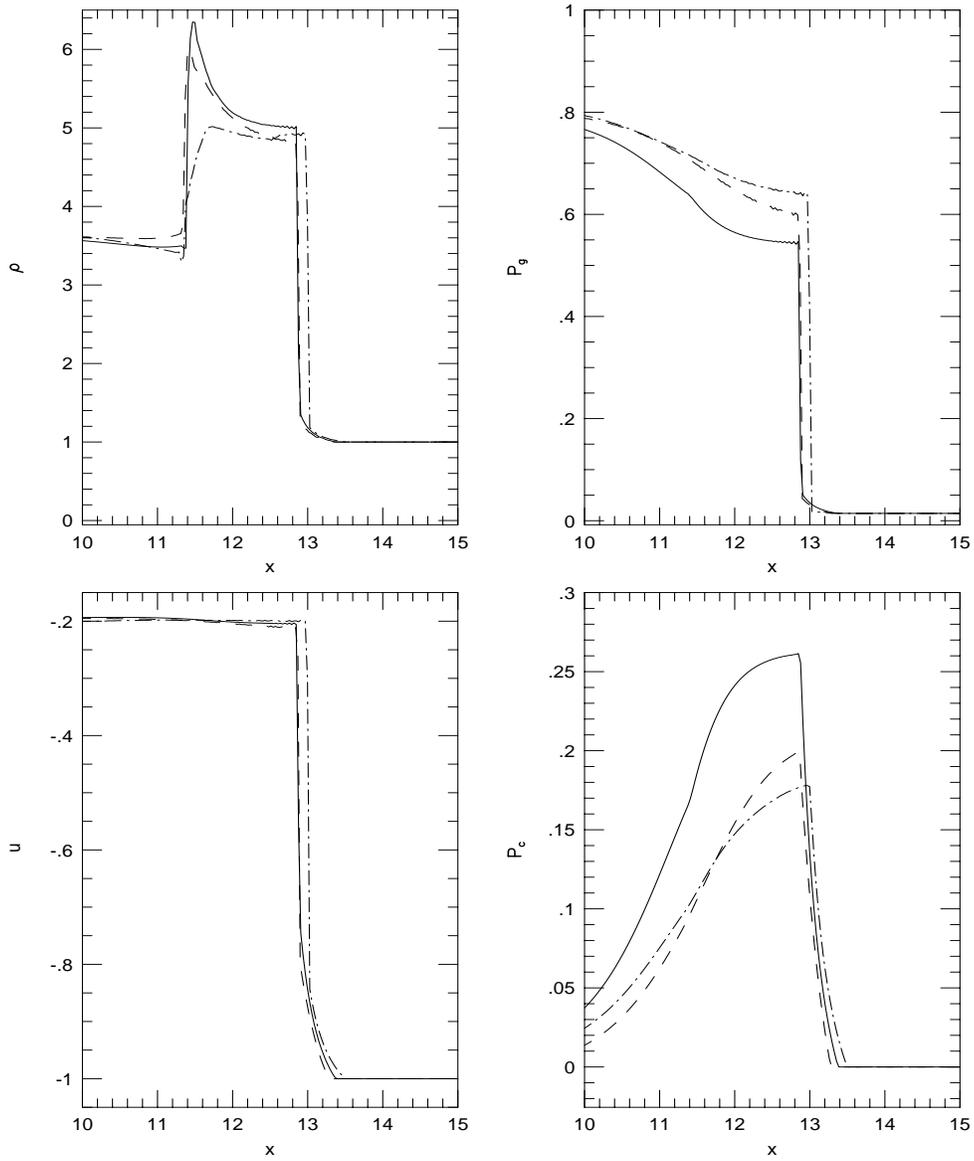

Fig. 5.— The flow structure of the model EMPd shock at $t = 8$ minutes. The line types are the same as Fig.2.

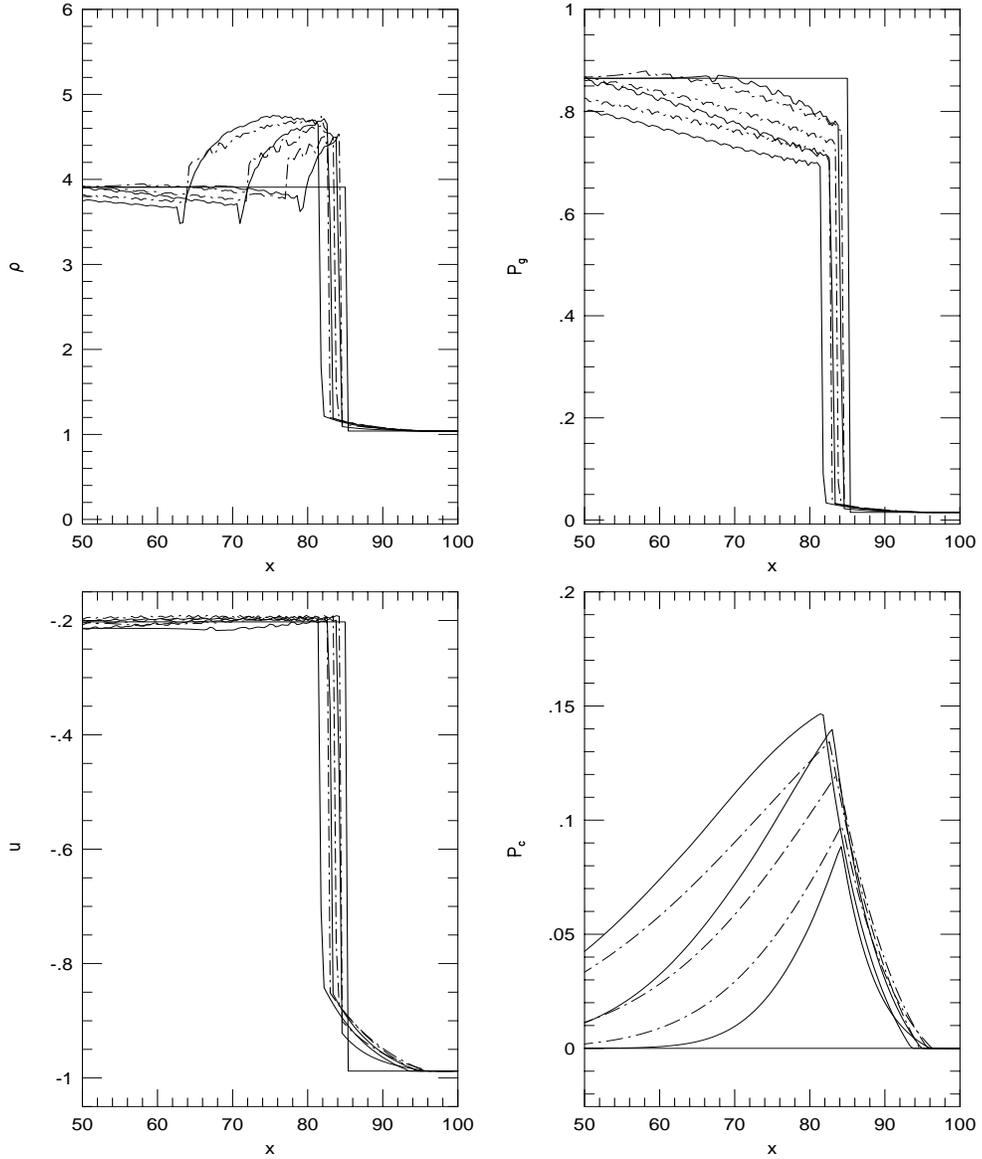

Fig. 6.— Time evolution of the EGBS shock as found from kinetic equation and two-fluid versions of the diffusive transport theory. This is the same shock shown in Fig. 3. Structures are shown at $t = 0$, 1, 4, 8 minutes. Line types are the same as those in Fig. 3.